\documentclass[english,prl,twocolumn]{revtex4}
\usepackage[T1]{fontenc}
\usepackage{amsmath,hyperref,amssymb}
\usepackage{graphicx,url}
\usepackage{float}
\usepackage{subfigure}
\newcommand{\eq}{\begin{equation}}
\newcommand{\feq}{\end{equation}}
\newcommand{\be}{\begin{equation}}
\newcommand{\ee}{\end{equation}}

\begin{document}

\title{Thermodynamics of spinning AdS$_4$ black holes in gauged supergravity}

\author{Chiara Toldo}

\affiliation{ Department of Physics, Columbia University, Pupin Hall, 538 West 120th Street\\
New York 10027, New York, USA }

\begin{center}
\begin{abstract} 
In this paper we study the thermodynamics of rotating black hole solutions arising from four-dimensional gauged $\mathcal{N}=2$ supergravity. We analyze two different supergravity models, characterized by prepotentials $F = -i X^0 X^1$ and $F= -2i \sqrt{X^0 (X^1)^3}$. The black hole configurations are supported by electromagnetic charges and scalar fields with different kinds of boundary conditions. We perform our analysis in the canonical ensemble, where we find a first order phase transition for a suitable range of charges and angular momentum. We perform the thermodynamic stability check on the configurations. Using the holographic dictionary we interpret the phase transition in terms of expectation values of operators in the dual field theory, which pertains to the class of ABJM theories living on a rotating Einstein universe. We extend the analysis to dyonic configurations as well. Lastly, we show the computation of the on shell action and mass via holographic renormalization techniques. 
\end{abstract}
\end{center}

\maketitle

\section{Introduction}\label{sect:intro}

It is a well known fact that Anti de Sitter (AdS) spacetime has a stabilizing effect on black hole thermodynamics. The gravitational potential of AdS spacetime acts as box of finite volume, making the total energy of the thermal radiation finite. Black holes can be in equilibrium with radiation at a fixed temperature, realizing a well defined canonical ensemble.

Thermodynamics of AdS black holes is particularly rich and has been extensively studied during the past years due to its relevance in applications of the AdS/CFT correspondence \cite{Maldacena:1997re}. In the simplest scenario, a thermal gas in AdS collapses into a Schwarzschild AdS black hole for a sufficiently large temperature \cite{Hawking:1982dh}. This process was interpreted via AdS/CFT as a confinement-deconfinement phase transition in the dual gauge theory \cite{Witten:1998zw}. More elaborate systems of black holes coupled to scalar and vector fields have been used in modeling phase transitions such as the superconductor one \cite{Hartnoll:2008vx}. 

The thermodynamics of static AdS Reissner-Nordstrom black holes is well known: in \cite{Chamblin:1999tk,Chamblin:1999hg} a first order phase transition was found among small and large AdS black holes. The phase diagram has interesting resemblance to the Van der Waals liquid-gas system. 

Spinning black holes have attracted, so far, less attention. The analysis of this paper makes some steps towards filling in this gap. The black holes we have observed in the sky have angular momentum and AdS rotating black holes provide a simplified scenario, an ideal playground to study the physics of such black holes. Indeed in Anti-de Sitter spacetime there exist Killing vectors describing rotating frames which have the property of being timelike everywhere,  in contrast to the asymptotically flat case. For this reason Kerr-AdS black holes can be in equilibrium with rotating thermal radiation all the way to infinity. We can study the thermodynamics of these configurations in an ensemble of fixed temperature, and relate it to that of the the boundary three-dimensional conformal field theory, which lives in a rotating universe. 

The
first studies of thermodynamics of rotating AdS black holes in four and five dimensions were carried out by \cite{Hawking:1998kw,Hawking:1999dp}, where the partition function of Kerr-AdS and Kerr-Newman AdS solutions was investigated and compared to that of a conformal field theory living on the boundary. Phase transitions of Kerr-Newman AdS black holes were investigated first by \cite{Caldarelli:1999xj} and subsequently the study was extended to higher dimensions, see for instance \cite{Altamirano:2013ane,Altamirano:2013uqa}. Further studies of the thermodynamics and the first law for rotating black holes in Anti-de Sitter were carried out in  \cite{Cvetic:2005zi,Gibbons:2004ai,Gibbons:2005vp}.

The analysis of this paper focusses on charged rotating AdS$_4$ black holes with scalar fields and represents an extension of the work initiated in \cite{Cvetic:1999ne}, and elaborated for instance in \cite{Hristov:2013sya}, regarding phase transitions of AdS black hole solutions of supergravity theories.

The black hole configurations we consider  are solutions of $\mathcal{N}=2$ Fayet-Iliopoulos (FI) $U(1)$-gauged supergravity in four dimensions (for a review, see \cite{Andrianopoli:1996cm}). We deal with solutions of two different models, characterized by prepotentials $F = -2i \sqrt{X^0 (X^1)^3}$ and  $F=- iX^0 X^1$. These models arise as  reductions of eleven-dimensional supergravity on $S^7$, as found in \cite{deWit:1981sst,deWit:1982bul} and \cite{Cvetic:1999au,Gauntlett:2001qs}. More details on the M-theory origin will follow later in the text. 

These rotating black holes were discovered respectively in  \cite{Chow:2010sf}  for $F = -2i \sqrt{X^0 (X^1)^3}$ model and recently in \cite{Chow:2013gba,Gnecchi:2013mja} for the  $F=- iX^0 X^1$ one. 
 They have spherical horizon topology and are characterized by electromagnetic charges, mass and angular momentum. The neutral scalar field supporting the configuration is genuinely complex and it depends both on the radial and angular coordinate $\theta$. This represents a new ingredient with respect to static solutions considered in \cite{Hristov:2013sya}, where the axion is consistently truncated and the dependence of the scalar field is only radial.  

Anticipating our final results, we find a first order phase transition between small and large spinning black holes when both the charges and the angular momentum are lowered under their critical values. This is reminiscent of what happens for rotating configurations without scalar fields, namely Kerr-Newman AdS black holes \cite{Caldarelli:1999xj}. 

Moreover, the asymptotic expansion of the scalar field at the boundary gives us information about the expectation value of the corresponding operators in the dual field theory, which belongs to the class of ABJM theories \cite{Aharony:2008ug} in a rotating Einstein universe. As we will explain later on, we find that the process can be described as a "liquid-gas like" phase transition. The corresponding ABJM-like dual theory is deformed by different operators depending on the boundary conditions satisfied by the scalar field, and this will be a distinguishing feature for the two different models.

\section{Rotating black hole solution, prepotential $F = -2i \sqrt{X^0 (X^1)^3}$}\label{sect:solutions1}

As a first example, we consider here the purely electric rotating AdS black holes of  \cite{Chow:2010sf}. They arise from $\mathcal{N}=2$ abelian FI (electrically) gauged four-dimensional supergravity with prepotential $F=-2i \sqrt{X^0 (X^1)^3}$. This theory is an abelian truncation of maximal $\mathcal{N} =8$ gauged supergravity (which arises as a reduction of M-theory on $S^7$ \cite{deWit:1981sst,deWit:1982bul}) obtained by retaining the $U(1)^4$ Cartan subgroup of $SO(8)$. In this model three of the four $U(1)$ gauge fields are identified.
 
The content of the theory consists of the gravity multiplet and one vector multiplet. After truncating the imaginary part of the complex scalar field, the bosonic Lagrangian reads  (see for instance \cite{Duff:1999gh,Toldo:2012ec}),
\begin{eqnarray}\label{LLLagr}
L&=&\frac{1}{2}R(g)-e^{\sqrt{6}\phi} F^0_{\mu\nu}F^{0\,\mu\nu}-3e^{-\sqrt{\frac{2}{3}}\phi} F^1_{\mu\nu}F^{1\,\mu\nu}\nonumber\\
&+&\frac{1}{2}\partial_\mu\phi \partial^\mu\phi +\large(g_0g_1e^{-\sqrt{\frac{2}{3}}\phi}+\frac{g_1^2}{3}e^{\sqrt{\frac{2}{3}}\phi}\large)\ .
\end{eqnarray}
In what follows we fix the value of the FI parameters to $g_0 = 1/\sqrt2$, $g_1= 3/\sqrt2$, hence we get the cosh-potential of \cite{Duff:1999gh} and the AdS radius $l$ is fixed to $l=1$. 
 
The rotating black hole configuration of \cite{Chow:2010sf} is supported by electric charge and purely real scalar. The imaginary part of the scalar is consistently set to zero, furthermore all components of the gauge field $A_1$ vanish. The rotating configuration with both vector fields and axions turned on remains unknown, while it is known in the static case \cite{Chow:2013gba}. Let us mention that, for static magnetic BPS solutions of this model, the microstate entropy counting was successfully achieved in \cite{Benini:2015eyy}. 

The black hole metric can be cast in the form
\eq
ds^2 = -f(dt + \omega_{y}dy)^2 + f^{-1}\left[v\left(\frac{dq^2}{Q} + \frac{dp^2}{P}\right) + PQ dy^2\right]\,,
\label{fibration}
\feq
with
\begin{eqnarray}
P (p) & = & (1-p^2)(j^2-p^2)  \,, \nonumber\\
Q (q)&=& q^2 + j^2 - 2mq + q^2(q^2 + 2ms^2 q + j^2)\,, \nonumber \\
                        \Delta_{\theta} & = & 1 - p^2\,, \nonumber \\
v & = & ((1-p^2)Q-V_r(j^2-p^2))(1-p^2) \,, \nonumber \\
V_r & = &\left(1+q^2\right) \left(2 m q s^2+ q^2+1\right)\,,\nonumber \\
f & = & \frac{v}{\sqrt{H} (p^2+q^2)} \,, \nonumber \\
\omega_y & = &\frac{2 c m P q \sqrt{1+j^2 s^2}}{\Xi v} \,,
\end{eqnarray}
and $\Xi  =  1- j^2 $. We defined $p=j \cos \theta$, $q$ is the radial coordinate, and $0 \leq \theta \leq \pi$, $0 \leq y <2 \pi$ is the usual range of Boyer-Lindquist coordinates. 
The dilaton has this form:
\eq
\phi = \frac12 \sqrt{\frac32} \log H\,,
\feq
where
\eq H = 1 +\frac{2 m q s^2}{\rho^2} \,, \qquad
\rho^2  =  q^2 +j^2 \cos^2 \theta \,,  \nonumber 
\feq
and $ s =  \sinh\delta$, $c = \cosh\delta$. The components of the non vanishing gauge field $A^0 $ read
\eq
A_t^0 = \frac{2 m q s c \Delta_{\theta}}{H \rho^2 \Xi}\,, \qquad A_{y}^0=  \frac{2 m q s  j \sqrt{1+j^2 s^2} \sin^2 \theta}{H \rho^2 \Xi}\,.
\feq
The solution admits a horizon and surrounding a ring-like singularity at $q=0$ for an appropriate choice of parameters $\delta,j,m$.  
This black hole configuration cannot be supersymmetric  for $j \neq 0$ \cite{Chow:2010sf}. In the static limit $j=0$ it reduces to the subset $A_1=0$ of the static electric AdS black holes found in \cite{Duff:1999gh}. The latter configurations admit a 1/2 BPS limit retrieved for $m \rightarrow 0$, $\delta \rightarrow \infty$. More details are in \cite{Duff:1999gh,Toldo:2012ec}.

The extremum of the scalar potential appearing in \eqref{LLLagr} is at $\phi_*=0$, which is the maximum. The mass of the scalar is $m_{\phi}^2=-2$, which lies in the interval
 \eq \label{bfrange}
 m_{BF}^2<m_{\phi}^2 <m_{BF}^2+1\,,
 \feq
where $m_{BF}$ is the Breitenlohner--Freedman mass \cite{Breitenlohner:1982jf}, which for our model is $ m_{BF}^2=-9/4$.  The asymptotic scalar field expansion\footnote{We shifted the radial coordinate in such a way that the metric expansion for $q \rightarrow \infty$ is of the form $g_{tt} = q^2/l^2+O(1)$, namely there is no linear term in $q$.} is
\begin{equation}\label{phi-asympt}
\phi=\frac{\alpha}{r}+\frac{\beta}{r^2}+{\cal{O}}(r^{-3})\ ,
\end{equation}
with 
\eq \beta =-\alpha^2/{\sqrt{6}} \qquad  and \qquad  \alpha= \sqrt{\frac32} m \sinh^2\delta\,.
\feq 
We have then $\beta=f(\alpha) = \kappa \alpha^2$ with a fixed value of $\kappa=-1/\sqrt6$. This kind of boundary conditions, called mixed boundary conditions, corresponds to a triple trace deformation of the boundary field theory \cite{Witten:2001ua,Hertog:2004ns,Papadimitriou:2007sj}. Mixed boundary conditions were previously considered in \cite{Hertog:2004dr}, where the authors show that there exist weakened boundary conditions on the metric and scalar fields which are compatible with asymptotic AdS  symmetries. Such boundary conditions allow in particular for a well-defined Hamiltonian \cite{Henneaux:2006hk}. Finally, in \cite{Hertog:2004bb} this study was first applied in the context of AdS black hole thermodynamics.

As a last remark, it is worth noting that the solution satisfy the same mixed boundary conditions as the static one. As we will see later, the boundary conditions on the scalar field will be different for the solution of the model $F= -i X^0 X^1$.  

\subsection{Thermodynamic quantities} 

The thermodynamic quantities for the solution described in the previous section have been computed in \cite{Chow:2010sf} and we give here a brief summary.
The area of the event horizon this gives us the entropy $S$:
\eq\label{entropyel}
S = \frac{\pi}{\Xi} \sqrt{(q_{\text h}^2 + j^2) (q_h^2 +j^2+2m s^2 q_h) }\,,
\feq
where $q_h$ is the radial location of the black hole horizon. The angular velocity of the horizon is
\eq 
\omega_{\text h}   = \frac{\sqrt{1+a^2 g^2 s^2} a (1+g^2 r_h^2)}{c (r_h^2+a^2)}\label{omega_hel}\,.
\feq
The temperature $T$ is:
\eq \label{TTel}
T= \frac{q_h^2 -j^2 + q_h^2 (3 q_h^2 +j^2 +4 m s^2 q_h)}{ 4 \pi q_h \sqrt{(q_h^2 +j^2)( q_h^2 +j^2 +2m s^2 q_h)} }\,.
\feq
One needs to use care in defining the mass for asymptotically Anti--de Sitter spacetimes. A rigorous method is the holographic renormalization procedure, which extracts the mass from the renormalized boundary stress energy tensor. We show explicitly the mass computation via holographic renormalization techniques in the last section of this paper. The mass $M$ is:
\eq
M =  m\frac{2+(1+j^2 ) s^2}{2 \Xi^2}\,.
\feq
This result agrees with the expression obtained with the Ashtekar-Magnon-Das (AMD) \cite{Ashtekar:1999jx,Ashtekar:1984zz} formalism, as computed in \cite{Chow:2010sf}. 
The angular momentum of the rotating black hole is computed as the Komar integral relative to the asymptotic Killing vector $k = \partial_{\phi} $, and it reads
\begin{equation}\label{Jel}
J =   -\frac1{8\pi }\oint_{\text{S}^2_{\infty}} dS^{\mu\nu}\nabla_{\mu}k_{\nu} =  m j c\frac{\sqrt{1+ j^2 s^2 }}{\Xi^2}\,.\end{equation}
The total electric charge is given by $Q_0$
\eq \label{elcharge}
Q_0 =  \frac{m s c}{ 2 \Xi}\,, \qquad Q_1=0\,,
\feq
and the electrostatic potentials assume this form:
\begin{eqnarray}
\Phi^0 = - \frac{2 m cs q_h}{q_h^2 +j^2 + 2m s^2 q_h}\,, \qquad \Phi^1=0\,.
\end{eqnarray}
The conserved charges we have defined in this section satisfy the first law of thermodynamics \cite{Chow:2010sf}
\eq \label{firstlaw}
dM = TdS + \Omega dJ - \Phi^{\Lambda} dQ_{\Lambda} + \chi_{\Lambda} d\pi^{\Lambda} \,,
\feq
where $\pi^{\Lambda}$ are the magnetic charges and $\chi_{\Lambda}$ are the associated magnetostatic potentials (which are set to zero in the solution described in this section). The quantity $\Omega$ appearing in \eqref{firstlaw} is defined as \eq \label{ome}
\Omega = \omega_{\text h} - \omega_{\infty}\,,
\feq
namely it is the difference between the angular velocity at the horizon and at infinity \cite{Caldarelli:1999xj}. Such angular velocity $\Omega$ act as chemical potential for the angular momentum of the fields in the dual field theory. For the solution described in this section, $\omega_{\infty} =0$, because the coordinates used are asymptotically static. Hence $\Omega = \omega_h$.

\subsection{Thermodynamics}

As we mentioned already,  Anti-de Sitter black holes can be in equilibrium with thermal radiation, therefore we choose an ensemble of fixed temperature. We will furthermore work first in an ensemble of fixed charge and angular momentum, namely we work in the canonical ensemble\footnote{Moreover, we will not consider the possible variation of the cosmological constant. For examples of analysis where this is done, refer to \cite{Cvetic:2010jb,Kubiznak:2012wp}}.  Our analysis will concern phase transitions between rotating single-centered configurations.

In order to see if a phase transition can arise, we need to investigate if there are multiple configurations with the same fixed value of temperature, charge and angular momentum. If there are different branches, the one characterized by lowest free energy will dominate the ensemble.
To see this we plot the temperature as a function of the entropy for fixed angular momentum and charges (Plot \ref{plot1}).
\begin{figure}[htb]
  \begin{center}\label{plot1}
    \includegraphics[scale=0.6]{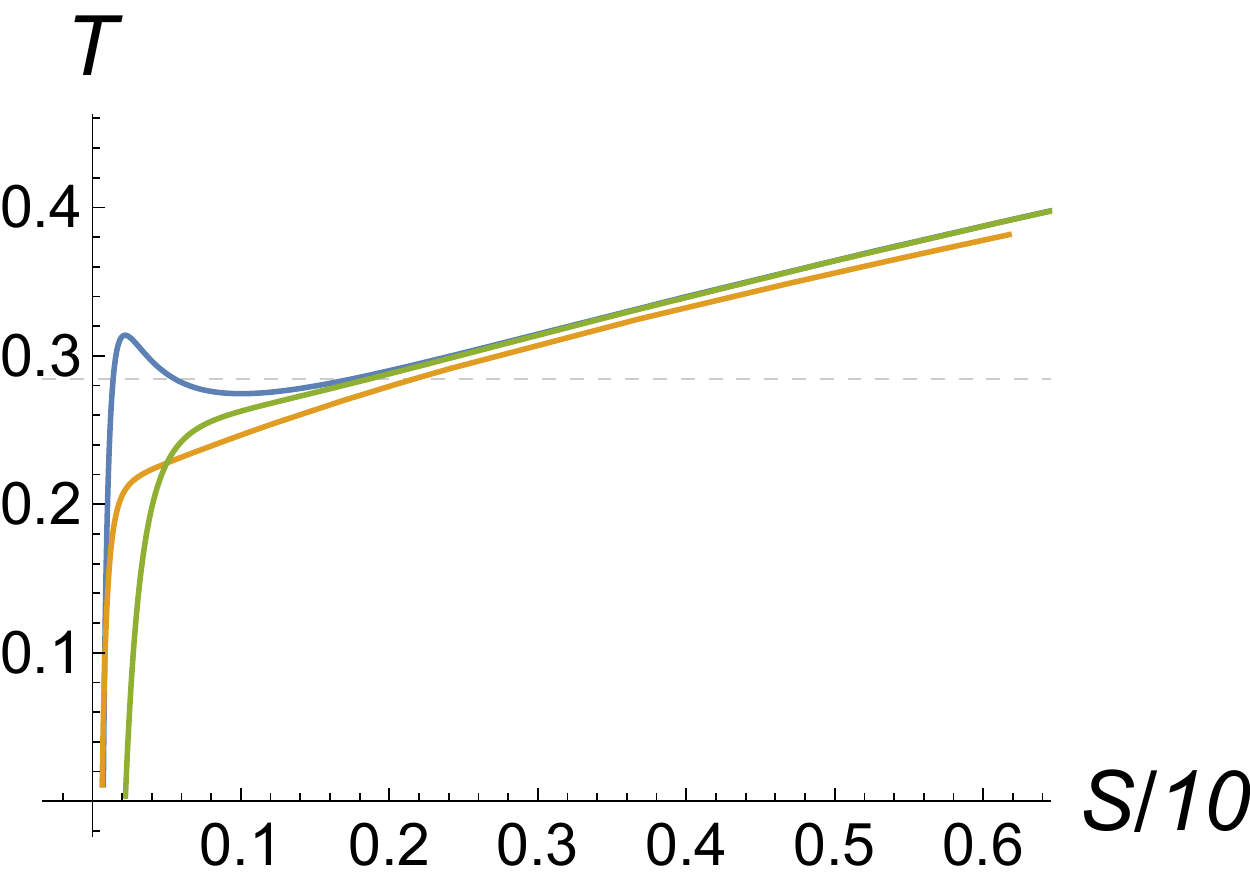}
      \caption{Behavior of the temperature $T$ plotted as a function of the entropy $S$ for the following sets of parameters: $J=-0.012$, $Q_0=0.004$ (blue), $J=-0.04$, $Q_0=0.004$ (green), $J=-0.012$, $Q_0=0.8$ (orange). \label{plot1}}
  \end{center}
  \end{figure}

\begin{figure}[htb]
  \begin{center}
    \includegraphics[scale=0.52]{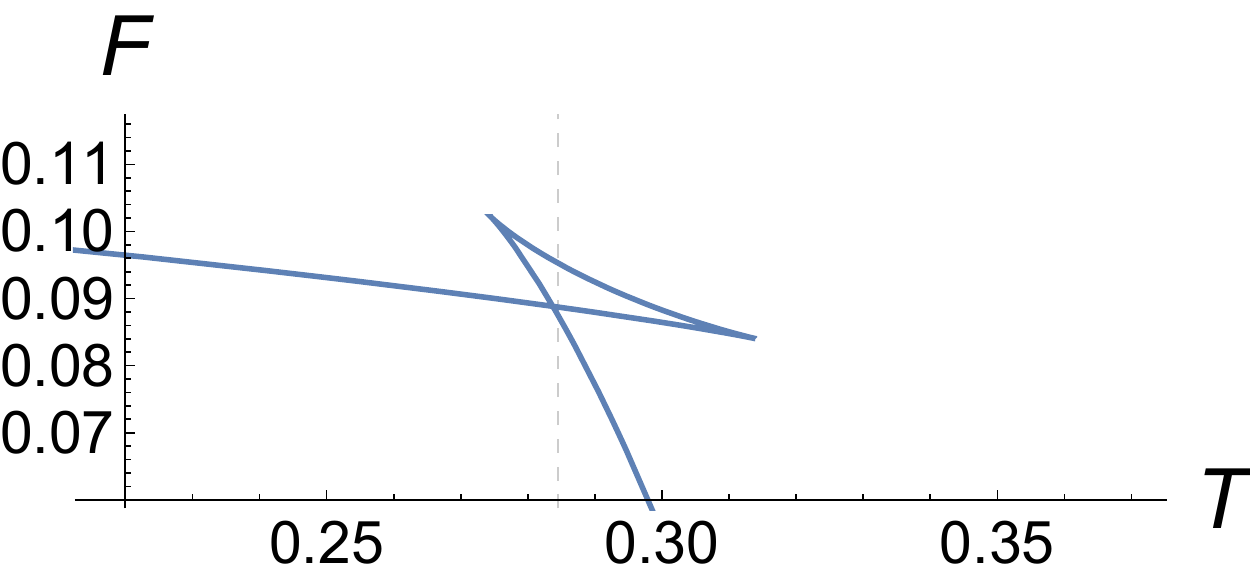}
  \end{center}
  \caption{Free energy plotted as a function of temperature for the following set of values: $J=-0.012$, $Q_0=0.004$, (blue line, which correspond to the blue line of the previous plot as well). The derivative of the free energy is discontinuous for $T \simeq 0.285$, resulting in a first order phase transition: a small rotating black hole becomes a large one.\label{fe1}}
\end{figure}

It turns out that there is a region of parameters $\{ Q_0, J \}$ for which three different black hole configurations with the same temperature, angular momentum and charge $T, J, Q_0 $ exist: see for example the blue line in  Fig. \ref{plot1}. As usual we denote them as small, medium sized and large black holes, depending on the value of the area of the event horizon. In this regime a first order phase transition arises:
 the plot of the Helmoltz free energy  $F$
\eq \label{helm}
F = M- TS
\feq 
in function of the temperature in Fig. \ref{fe1}  shows a discontinuity in the first derivative around $T = 0.285$. 
For the parameters corresponding to the orange and green lines in Fig. \ref{plot1} the free energy is instead a monotonically decreasing smooth function of $T$, hence no transition occurs.

The process involves small black holes turning into large ones upon increasing the temperature $T$. Medium sized black holes have always higher free energy, therefore they never dominate the ensemble. We find an upper bound for the angular momentum in order for the phase transition to take place:
\eq
 |J_{c}| \simeq 0.0235 \,.
\feq
No phase transition arises if $J>J_{c}$. If instead $J$ is lowered below $J_c$ there is a range of charge $Q_0$ for which the phase transition arises. Sampling the space of parameters, we found however no upper bound on the value of $Q_0$, namely even for large values of $Q_0$ a phase transition takes place if we sufficiently lower the angular momentum\footnote{We verified this up to values of charges $Q_0 = 350$.}.
This is somewhat different with respect to what happens in the case of the Kerr-Newman AdS black hole studied in  \cite{Caldarelli:1999xj}. An upper bound for both charge and angular momentum was found there.

Sampling numerically the space of parameters, we could see that there are no multiple configurations for $T=0$, hence in this ensemble there is no quantum phase transition among extremal black holes. Furthermore, we fixed here the AdS radius to $l=1$ in order to simplify the computations, but one can ask what is the effect of the variation of the cosmological constant on the critical value of the charges and angular momentum. As found in \cite{Chamblin:1999tk}, an increase of the cosmological constant corresponds to a decrease in the critical value of the charge. We expect that the same considerations hold for configurations with scalars, and it would be interesting to verify this in our case.

\begin{figure}[htb]
  \begin{center}
    \includegraphics[scale=0.54]{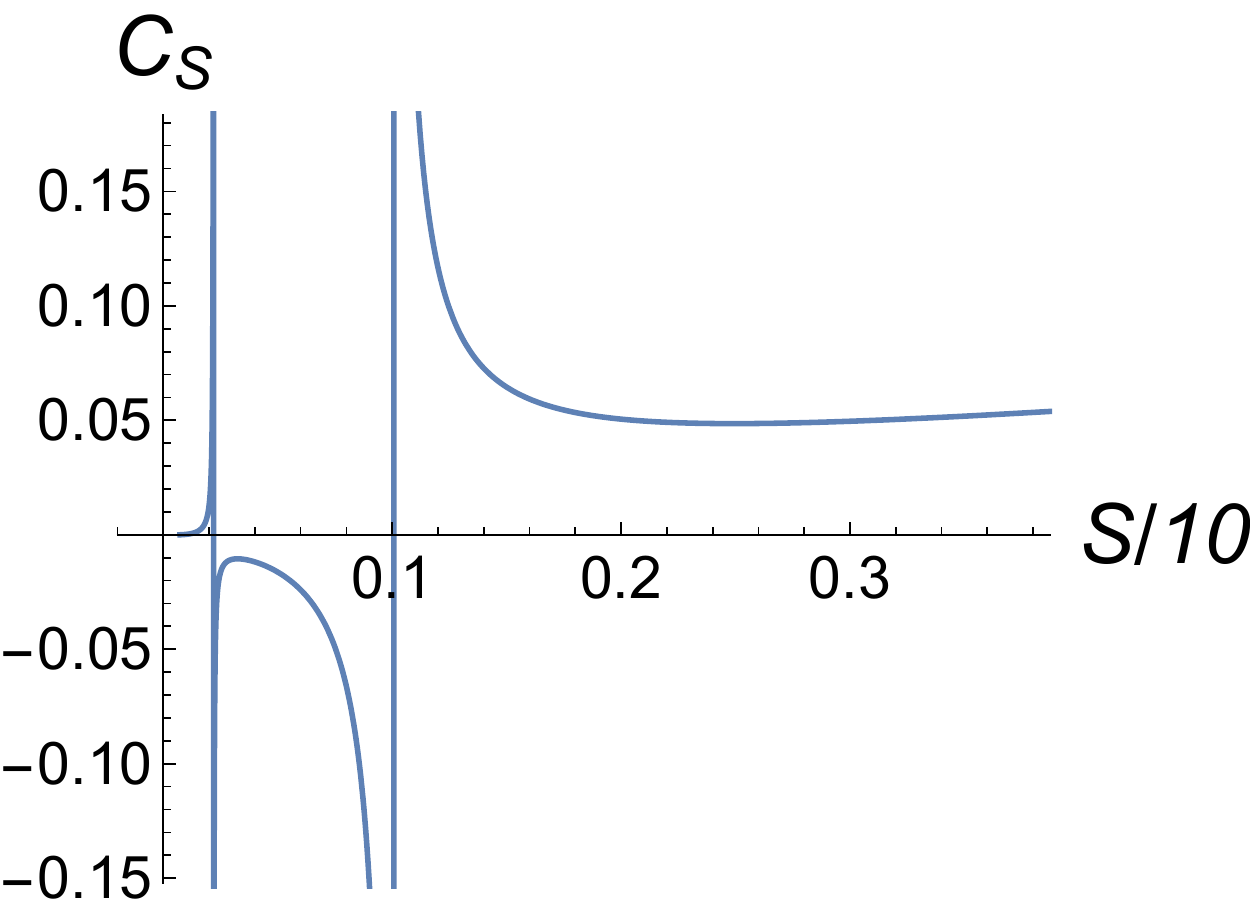}
  \end{center}
  \caption{Specific heat for the three branches of solutions existing for the parameters $Q_0=0.004$, $J=-0.04$, plotted as a function of the entropy $s=S/10$. Comparing this graphs with Fig. \ref{plot1}, one can see that the small black holes ($s< 0.02$) and the large ones ($s>0.1$) have positive specific heat, while for the medium ones $C_S$ is negative. At the turning points the specific heat diverges.\label{spec_heat}}
\end{figure}

We furthermore checked the thermodynamical stability of the various black hole branches by computing the specific heat at fixed charge and angular momentum
\eq \label{sph}
C_S = T \, \left( \frac{\partial S}{ \partial T} \right)_{Q,J} \,.
\feq
Medium sized black holes have negative specific heat and hence they are unstable. As expected, small black holes and large ones are always stable (see Fig. \ref{spec_heat}). 

Let us conclude this section with two observations. In the process we have described the area of the event horizon increases. On the other hand,  one can also verify that the angular velocity at the horizon $\omega_h$ decreases through the phase transition. This is reminiscent of what happens in rigid bodies due to conservation of angular momentum: the angular velocity decreases upon increasing the moment of inertia of the black hole.

Secondly, there is one caveat to our analysis. We have investigated the thermodynamical stability of the configurations, but a more thorough investigation of their stability (e.g. mechanical stability) is needed.  As an example, let us mention the question of superradiance for black holes with angular momentum (see for example \cite{Brito:2015oca} and references therein). Indeed, in cosmological Einstein-Maxwell theories it was proven that small rotating AdS black holes suffer from superradiance \cite{Cardoso:2004hs,Aliev:2015wla,Delice:2015zga}, hence they are unstable. Including the analysis of superradiant instability (along the lines for example of \cite{Aliev:2008yk,Birkandan:2015yda}) would be important in order to understand the true nature of the phase transition here discovered. This analysis is beyond the scope of this work and we leave it as a future direction. 

\subsection{Dual field theory interpretation}

As we mentioned already, the model comes from reduction of M-theory on $S^7$. Explicit uplifts for static black holes were constructed in \cite{Cvetic:1999xp} and the uplifted configurations are interpreted in eleven dimensions as the decoupling limit of spinning M2-branes.

We now interpret the phase transition we have found in the gravity side in the dual field theory, which is a three dimensional superconformal Chern-Simons field theory belonging to the class of ABJM models  \cite{Aharony:2008ug}, in presence of non vanishing chemical potentials for electric charges and angular momenta. Due to the presence of mixed boundary conditions $\beta = \kappa \alpha^2 $ on the scalar field, the ABJM action $S_{ABJM}$ is characterized by a triple trace deformation of the form
\eq
S=S_{ABJM}+ \kappa \int {\cal O}_1^3\,,
\feq
where $\mathcal{O}_1$ is an operator of order one \cite{Witten:2001ua,Berkooz:2002ug,Hertog:2004ns}. The deformation is therefore marginal. Example of such operators are bilinears of boundary scalars $\Phi$ transforming under the global R-symmetry group, ${\cal O}_1={\rm Tr}(\Phi^I a_{IJ}\Phi^J)$, where $a$ is a constant matrix \cite{Hertog:2004ns}. In our case, the parameter $\kappa$ is fixed to be $\kappa=-1/{\sqrt 6}$ for all configurations.

\begin{figure}[htb]
  \begin{center}
      \includegraphics[scale=0.53]{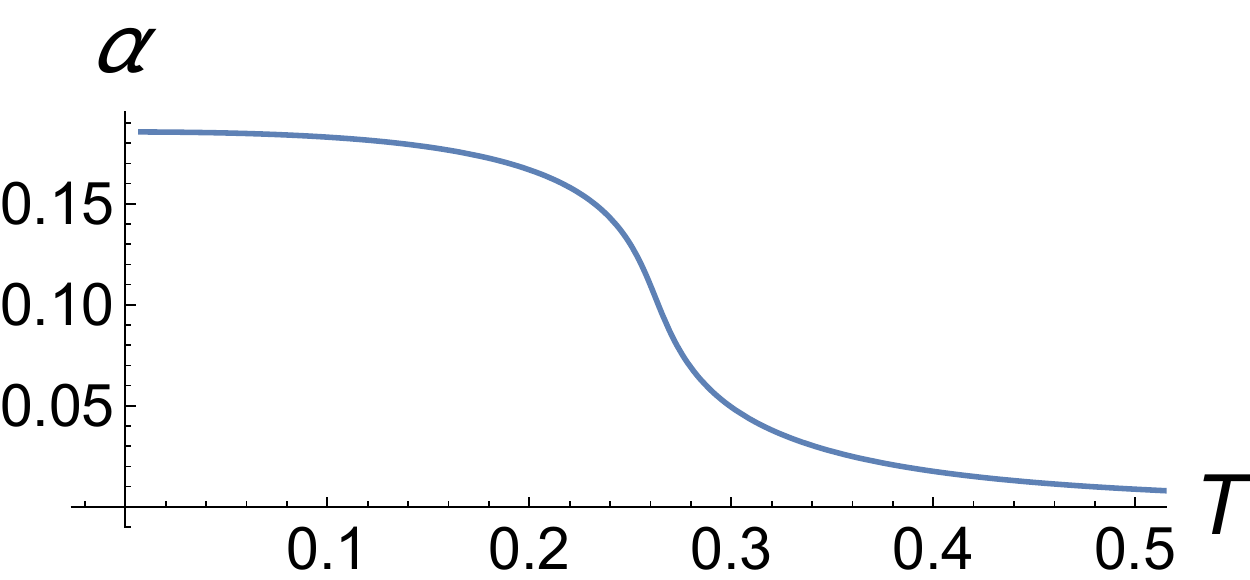}
    \includegraphics[scale=0.53]{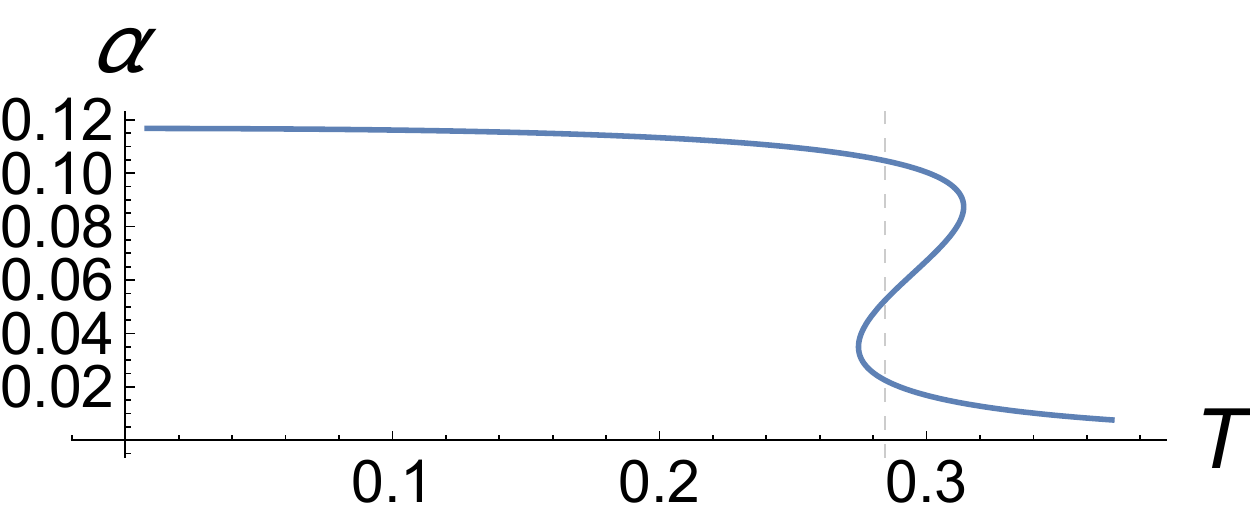}
  \end{center}
  \caption{Behavior of $\alpha$ for the set of parameters $Q_0=0.004$ and $J=-0.04$ (above) and $Q_0=0.004$ and $J=-0.0012$ (below). In the first case (top), we are outside the critical region hence there is no phase transition: $\alpha$ is a monotonic decreasing function of the temperature $T$. In the second case (bottom) the parameter $\alpha$ exhibits a "wiggle", typical of a liquid-gas like phase transition. The dashed line denotes the temperature at which the phase transition arises. We notice here that Maxwell's construction for determining the temperature of the phase transition does not hold. Indeed there are examples where the area bounded by the dashed line and the blue curve on the right-hand side is not equal to that on the left hand side.\label{cus2}}
\end{figure}

The holographic dictionary in presence of mixed boundary conditions was developed in \cite{Papadimitriou:2007sj}, where it was found that mixed boundary conditions enforce a multi-trace deformation of the field theory dual to the Neumann boundary conditions. The source of the deformation is set to zero for every solution satisfying $\beta = \kappa \alpha^2$, and $\alpha$ is the expectation value of the operator $\mathcal{O}_1$,  
$
\langle {\cal O}_1 \rangle =\alpha\,
$.
We therefore take $\alpha$ as order parameter for the phase transition, and we plot it as a function of the temperature in Fig. \ref{cus2}. In analogy with \cite{Hristov:2013sya} there is a different behaviour in the two different regimes above the critical values of the charges and angular momentum, and below. Below the critical values the phase transition manifests itself in the dual field theory as a liquid/vapor -like transition, like what happens for static magnetic black holes \cite{Hristov:2013sya}. 

Since the expectation value $\alpha$ is never zero, R-symmetry is always broken, for any finite temperature. 
This is in contrast with the situation of the holographic superconductor phase transition: in the superconductor case the operator ${\cal O}_1$ acquires a nonzero expectation value only below a certain temperature and the phase transition manifests itself as spontaneous breaking of a global $U(1)$ symmetry.

\section{Rotating black hole solution of the prepotential $F= -i X^0 X^1$}\label{sect:solutions}
We consider the rotating purely magnetic AdS black holes of \cite{Gnecchi:2013mja} arising from $\mathcal{N}=2$ abelian FI (electrically) gauged supergravity with prepotential $F=-i X^0 X^1$. This model is a truncation of $\mathcal{N}=4$ $SO(4)$ gauged four-dimensional supergravity, which was obtained by Kaluza-Klein reduction of $D=11$ supergravity on $S^7$ in \cite{Cvetic:1999au,Gauntlett:2001qs}. In the static limit, these configurations arise as M-theory membranes wrapping holomorphic curves of Calabi--Yau threefolds \cite{Gauntlett:2001qs}.

Again we consider the gravity multiplet and one vector multiplet. The bosonic Lagrangian is:
\eq\label{act}\nonumber
S=\int {\rm d}^4x\, \sqrt{-g} \bigg[ \frac12 R - \frac{ \partial^{\mu} \tau \partial_{\mu} \bar{\tau}} {(\tau + \bar\tau)^{2}}+ I_{\Lambda \Sigma}(\tau, \bar{\tau}) F_{\mu \nu}^{\Lambda} F^{\mu \nu |\Sigma} \, +
\feq
\eq
+ \frac12 R_{\Lambda \Sigma} (\tau, \bar{\tau}) \epsilon^{\mu \nu \rho \sigma} F_{\mu \nu}^{\Lambda} F_{\rho \sigma}^{\Sigma} - V(\tau, \bar{\tau})\bigg] \, ,
\feq
where the complex scalar
is denoted by $\tau$. The vector kinetic matrix is of this form:
\eq
{\cal N} = \left(\begin{array}{cc} -i\tau & 0 \\ 0 & -\frac i\tau\end{array}\right)\ ,
\feq
and it requires ${\mathrm{Re}}\tau>0$. The functions $R_{\Lambda \Sigma}$ and $I_{\Lambda \Sigma}$ appearing in the action \eqref{act} are respectively the real and imaginary part of the kinetic matrix $\mathcal{N}$.

Given the electric gauging with FI parameters $g_0$ and $g_1$, the scalar potential is
\eq
V = -\frac4{(\tau+\bar\tau)}(g_0^2 + 2g_0g_1 (\tau + \bar\tau)
+ g_1^2\tau\bar\tau)\ , \label{pot_su11}
\feq
which admits a maximum at the value $\tau=\bar\tau=|g_0/g_1|$.

The black hole solution is characterized by this metric:
\begin{eqnarray}\label{met}
ds^2 &= & -\frac{Q(q)}{(q^2-\Delta^2+j^2\cos^2\!\theta)}\left[dt + \frac{j\sin^2\!\theta}{\Xi} d\phi\right]^2  \\
&+ & (q^2-\Delta^2+j^2\cos^2\!\theta)\left(\frac{dq^2}{Q(q)} + \frac{d\theta^2}{\Delta_{\theta}}\right) +\nonumber \\
             & + &\, \frac{\Delta_{\theta}\sin^2\!\theta}{(q^2-\Delta^2+j^2\cos^2\!\theta)}\left[jdt +
                   \frac{q^2+j^2-\Delta^2}{\Xi}d\phi\right]^2\,, \label{metr-nonextr-spher-X0X1} \nonumber
\end{eqnarray}
where $0 \leq \theta \leq \pi$ and $0 \leq \phi <2 \pi$. The functions $\Delta_{\theta}$ and $\Xi$ are defined as
\begin{equation}\label{deltatheta}
\Delta_{\theta} = 1 - \frac{j^2}{l^2}\cos^2\!\theta\,, \qquad  \Xi = 1- \frac{j^2}{l^2}\,.
\end{equation}
The real parameters $j$ and $\Delta$ are related respectively to the angular momentum and the mass (alternatively, the scalar hair) of the black hole, in a way that we will soon specify. The parameter $l$ is the AdS radius. The gauge fields are magnetic
\eq
A^\Lambda = \frac{\mathsf{P}^\Lambda(dt + q_1q_2dy)}W j \cos \theta\,, \qquad \Lambda =0,1 \,,\feq
where 
\eq
W = q_1q_2 +j^2\cos^2\theta\,,  \quad q_1 = q-\Delta\,, \quad q_2=q+\Delta\,.
\feq 
The constants $\mathsf{P}^\Lambda$ are proportional to the magnetic charges $\pi^{\Lambda}$:
\eq \label{magnch}
\pi^\Lambda = \frac1{4\pi}\oint_{\text{S}^2_{\infty}} F^\Lambda = -\frac{\mathsf{P}^\Lambda}{\Xi}\,.
\feq
The polynomial $Q(q)$ appearing in \eqref{met} is given by
\eq 
Q = a_0 + a_1q + a_2q^2 + a_4q^4\,, \label{QP}
\feq
with:
\begin{equation} \nonumber
a_0 = (j^2-\Delta^2)\left(1-\frac{\Delta^2}{l^2}\right) + 2l^2\left(g_0^2{\mathsf{P}^0}^2 +
g_1^2{\mathsf{P}^1}^2\right)\,, 
\feq
\eq
 a_1= \frac{2 l^2 (g_0^2{\mathsf{P}^0}^2 - g_1^2{\mathsf{P}^1}^2)}
{ \Delta}\,, \quad
a_2 = 1-\frac{\Delta^2}{l^2} + \frac{j^2-\Delta^2}{l^2}\,, 
\nonumber
\feq
\eq
 a_4 = 1/l^2 \equiv 4g_0g_1 \,.
\feq

The scalar field $\tau$ is complex, with both non vanishing imaginary and real part, and with nontrivial dependence on the radial and angular coordinates $q$ and $\theta$:
\eq
\tau = 
\frac{X^1}{X^0} = \frac{g_0}{g_1} \frac{ W+i 2 \Delta j \cos \theta}{q_1^2+j^2 \cos^2 \theta}\,, \label{ansatz-A-tau-magn}
\feq

The solution in total depends on the two FI parameters $g_0$ and $g_1$, which we consider as input parameters of the theory, and on the four parameters ${\mathsf{P}^1},{\mathsf{P}^0},j,\Delta$ representing the conserved charges associated with the black hole: two magnetic charges, angular momentum and mass.

Notice that for $g_0\mathsf{P}^0=g_1\mathsf{P}^1$ and $\Delta=0$, the parameter $a_1$ can be arbitrary: the scalar field is constant and the solution reduces to the Kerr-Newman-AdS solution with magnetic charge. On the other hand, for zero rotation parameter, $j=0$,
\eqref{metr-nonextr-spher-X0X1} boils down to the static nonextremal black holes with
running scalar constructed in \cite{Klemm:2012yg}. Furthermore, the BPS rotating solution \cite{Klemm:2011xw}  is recovered for 
\eq
\mathsf{P}^{\Lambda} = \frac{1}{4g^{\Lambda}} \left( 1+\frac{j^2}{l^2} \right)\,,
\feq
whose static limit \cite{Cacciatori:2009iz} can be found by setting $j=0$. Both static and stationary BPS limits correspond to naked singularities. 

The solution presented here admits a horizon shielding the singularity for a suitable choice of parameters, and in that case it represents a rotating AdS black hole. The location of the horizon is given by the largest root of the quartic equation $Q(q)=0$. The location of the singularity depends on $\theta$\footnote{This is different than the Kerr-Newman black hole case. The latter configuration is recovered for $\Delta=0$, therefore the only possible solution to eq. \eqref{sing} is $\theta = \pi/2 $ and $q=0$. In other words, the singularity of the Kerr-Newman solution is located at the equatorial plane and it is a ring. The solution \eqref{met} instead has a surface-like singularity, whose radius varies with $\theta$.}: it is located at
\eq \label{sing}
 (q^2-\Delta^2+j^2\cos^2\!\theta) =0 \,.
\feq
We recall here that the presence of the scalar field  accounts for the parameter $\Delta$ and the (constant scalars) Kerr-Newman AdS solution is retrieved for $\Delta=0$.

Before concluding this section, with an eye on the AdS/CFT description, let us look at the boundary. Taking now the asymptotic form of eq. \eqref{fibration}, we see that the boundary metric approaches the form
\eq \label{as_bound}
ds^2= \frac{q^2 \Delta_{\theta}}{ \Xi} \left[-dt^2 +\frac{\Xi d \theta^2}{\Delta_{\theta}^2}- \frac{\sin^2 \theta}{\Delta_{\theta}} \left( d \phi + \frac{j}{l^2} dt\right)^2  \right]\,.
\feq
This is not the
standard metric on $R \times S^2$, due to the fact that there is a non-zero angular velocity at infinity $\omega_{\infty}$. The metric in square bracket is that the Einstein space $R \times S^2$ seen by a rotating frame of reference. 
The coordinate change
\eq
\phi' = \phi + \frac{j}{l^2}t \,, \qquad \Xi \tan^2\theta' = \tan^2 \theta
\feq
brings the boundary in the standard form $R \times S^2$ up to the conformal factor $ \Xi \cos^2 \theta/ \cos^2 \theta'$. Hence the boundary metric in \eqref{as_bound} is conformal to the standard boundary of four-dimensional AdS space. More details can be found for instance in  \cite{Gibbons:2004ai,Gibbons:2005vp,Papadimitriou:2005ii}.

\subsection{Thermodynamics quantities and scalar asymptotics}

 For simplicity from now on we fix the FI parameters $g_0=g_1 =1/2$, so that the AdS radius is one. Our solution is supported by a complex scalar field  $\tau$ and the scalar potential \eqref{pot_su11}  of the model admits a maximum for 
\be
\tau=\bar\tau=1\,.
\ee
At this extremum the value of the potential is $V^* = -3=  \Lambda$, where $\Lambda$ is the cosmological constant.
We define
\eq \label{param_field}
\tau = e^{\sqrt2 (\varphi +i \chi) }  \,, \qquad \bar\tau = e^{\sqrt2 (\varphi -i \chi) } \,.
\feq
The fields $\chi$ and $\varphi$ are eigenmodes of the Hessian, and their mass is
\eq
m_{\chi}^2 = m_{\varphi}^2= -2\,.
\feq
Once again the mass eingevalues lie both in the interval
$-\frac94 < m^2 < -\frac94 +1 $ allowing for both choices of quantization. 

We have the following asymptotic boundary expansion\footnote{In these coordinates the metric expansion for $q \rightarrow \infty$ is already of the form $g_{tt} = q^2/l^2+O(1)$.} ($q \rightarrow \infty$) for $\varphi$  :
\eq \label{as1}
\varphi = \frac{\alpha}{q} + \frac{\beta}{q^2} + O(q^{-3})
\feq
where 
\eq \label{bc1}
\alpha =  \sqrt2 \Delta \,, \qquad
 \beta=0 \,.
\feq
The parameters $\alpha$ and $\beta$ are independent of the angular coordinate $\theta$: the dependence of the scalar field on $\theta$ starts at order $q^{-3}$. 
The second eigenmode $\chi$ has the following boundary expansion: 
\eq \label{as2}
\chi = \frac{\iota}{q}+  \frac{  \gamma}{q^2} + O(q^{-3})\,, 
\feq
\eq \label{bc2}
\iota=0 \,, \qquad
\gamma =  \sqrt2 \Delta j  \cos \theta \,.
\feq
We see that there is no order $q^{-1}$ mode, and the coefficient $\gamma$ depends explicitly on the variable $\theta$. Anticipating the discussion of the next sections, we see that there is no source term in this case, but nevertheless there is a nonzero expectation value for an operator of dimension two. The latter depends on the angular variable and therefore spontaneously  breaks rotational invariance in the boundary.

Let us now discuss the thermodynamic quantities. We refer to \cite{Gnecchi:2013mja} for details. Denoting once again the radial coordinate of the event horizon by $q_h$, entropy and temperature read
\eq\label{entropy}
S = \frac{\pi}{\Xi}(q_{\text h}^2 + j^2 - \Delta^2)\,, \quad T= \frac{Q'(q_h)}{ 4 \pi  (q_h^2+j^2-\Delta^2)}\,,
\feq
while the angular velocities of the horizon $\omega_{\text h}$ and at infinity $\omega_{\infty}$ are respectively
\eq 
\omega_{\text h} = -\frac{j\Xi}{q_{\text h}^2 + j^2 - \Delta^2}\,, \label{omega_h}
\qquad
\omega_{\infty} = \frac j{l^2}\,.
\feq
The mass $M$ and the angular momentum have the following expressions
\eq
M = -\frac{a_1}{2\Xi^2}\,, \qquad J = -j M\,.
\feq
The magnetic charges are defined in \eqref{magnch}
and the magnetostatic potentials assume this form \cite{Gnecchi:2013mja}:
\eq
\chi_\Sigma = \frac{\pi}{4SM}\pi^0\pi^1\eta_{\Sigma \Pi}\pi^\Pi + \left(\frac{l^2}{M} +
\frac{S}{\pi M}\right)g_{\Sigma}^2\pi^{\Sigma}\,, \label{magn-pot} \nonumber
\feq
where the matrix 
$
\eta = \left(\begin{array}{cc} 0 & 1 \\ 1 & 0 \end{array}\right)\,,
$
and there is no summation over $\Sigma$ in the last term. These quantities satisfy the Christodoulou-Ruffini mass formula as explained in \cite{Gnecchi:2013mja}.

The conserved charges we have defined in this section satisfy the first law of thermodynamics \eqref{firstlaw}. Moreover, we verified that they also satisfy the Smarr relation, which will be useful later in simplifying the computation of the free energy. The Smarr relation reads
\eq \label{smarr}
M = 2 TS + 2\Omega J + \chi_{\Sigma} \pi^{\Sigma} - 2 \Lambda \Theta_{tv}
\feq
where the cosmological constant $\Lambda$ is defined as $\Lambda = -3/ l^2$, and
\begin{eqnarray}
\Theta_{tv} & = &  -\frac{S}{ 2 M \pi^2}   \left(\frac{g_0 {(\pi^0)}^2}{g_1}+\frac{g_1 (\pi^1)^2}{g_0}\right) \\
&- &\frac{1}{6M } \left(J^2+\frac{S^2}{2 \pi ^2}\right)+\frac{\Lambda S^3}{36 M \pi ^3}\,. 
\end{eqnarray}
The quantity $\Theta_{tv}$ is called "thermodynamic volume" and it corresponds to the spatial volume inside the black hole horizon \cite{Kastor:2009wy,Cvetic:2010jb}. The cosmological constant acts as a pressure on the system and it is the variable conjugate to $\Theta_{tv}$. We will not consider variations of $\Lambda$ in the first law, which for us reads as is \eqref{firstlaw}.

\subsection{Thermodynamics and phase transitions}

We choose to analyze the thermodynamics of the black solutions described above in the canonical ensemble. The supergravity model is characterized by electric gauging, hence the fermions of the theory are electrically charged under the gauge group $U(1) \times U(1)$. The total magnetic charges $\pi^0$, $\pi^1$ of the black hole should then remain fixed, unless we assume the presence of other solitonic solutions that could carry away the magnetic charge from the black hole, or of multiple center configurations. We will not take this possibility into consideration and our analysis will concern phase transitions between rotating single-centered configurations. As we did in the electric case, we keep the total angular momentum $J$ fixed as well. 

We provide plots for various sets of parameters\footnote{We traded the dependence in $\Delta$ for the one in $j$ by means of the relation
$
J = \frac{a_1j}{2\Xi^2}\,.
$}.
 $J$, $\pi^0 $ and $\pi^{1}$ in Fig. \ref{cup0}. Similar to the electric rotating black holes analyzed previously, for a certain set of charges $\{ \pi^0, \pi^1, J \}$ three different black holes coexist (the blue line in  Fig. \ref{cup0}), and the plot of the free energy $F$ as \eqref{helm} in function of the temperature in Fig. \ref{fe}  shows again a first order phase transition between small and large configurations.

\begin{figure}[htb]
  \begin{center}
    \includegraphics[scale=0.55]{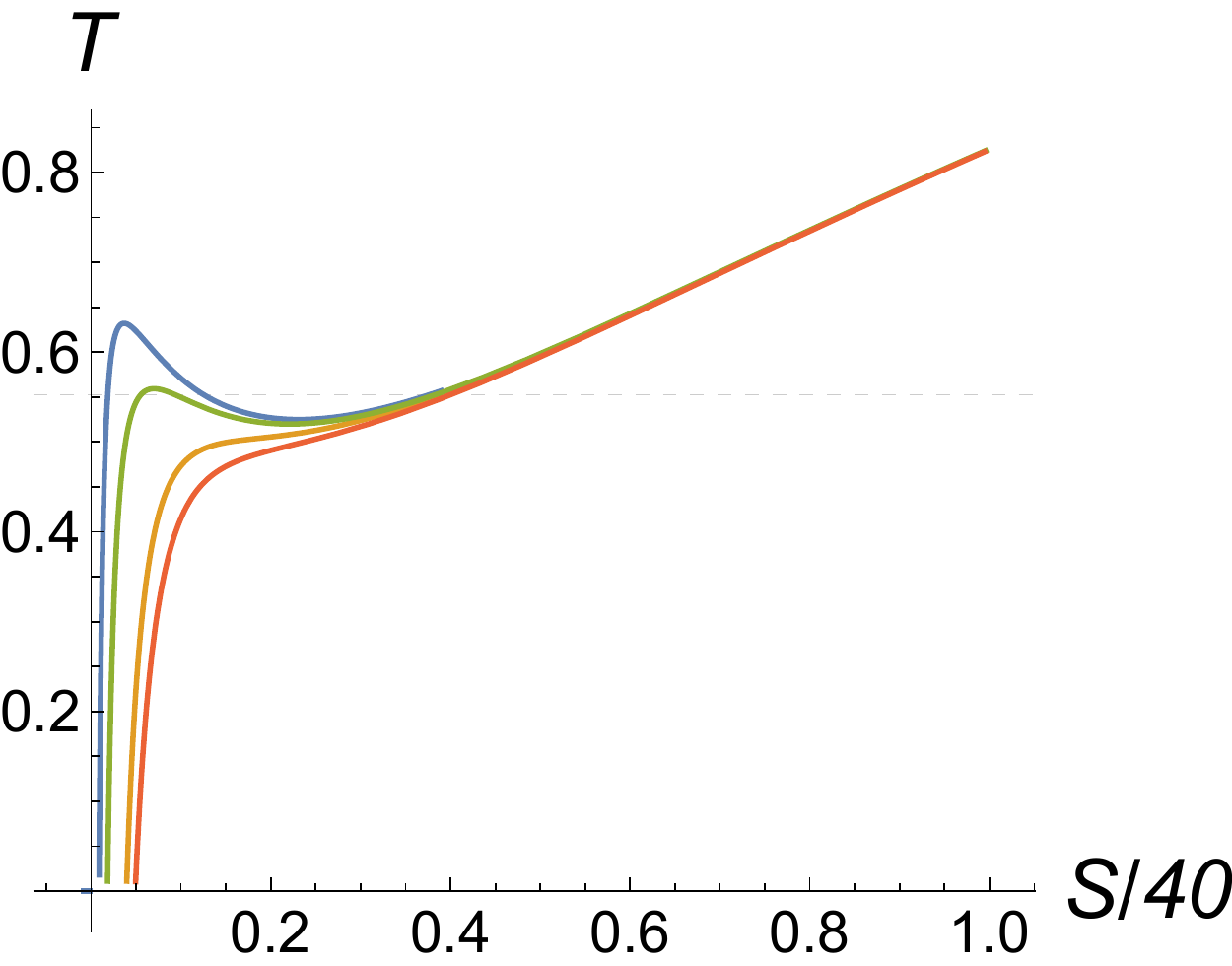}
  \end{center}
  \caption{Behavior of the temperature $T$ plotted as a function of the entropy $S$. For simplicity we fixed the value of $\pi^0 =10$ and we plotted for the following sets of conserved charges: $J=-1$, $\pi^0=10$, $\pi^1=0.05$ (blue), $J=-1$, $\pi^0=10$, $\pi^1=0.5$ (orange), $J=-2.2$, $\pi^0=10$, $\pi^1=0.05$ (green), $J=-2.2$, $\pi^0=10$, $\pi^1=0.6$ (red). We later investigate further sets of parameters as well, and we give the range of values for the occurrence of the phase transition. \label{cup0}}
\end{figure}

\begin{figure}[htb]
  \begin{center}
    \includegraphics[scale=0.48]{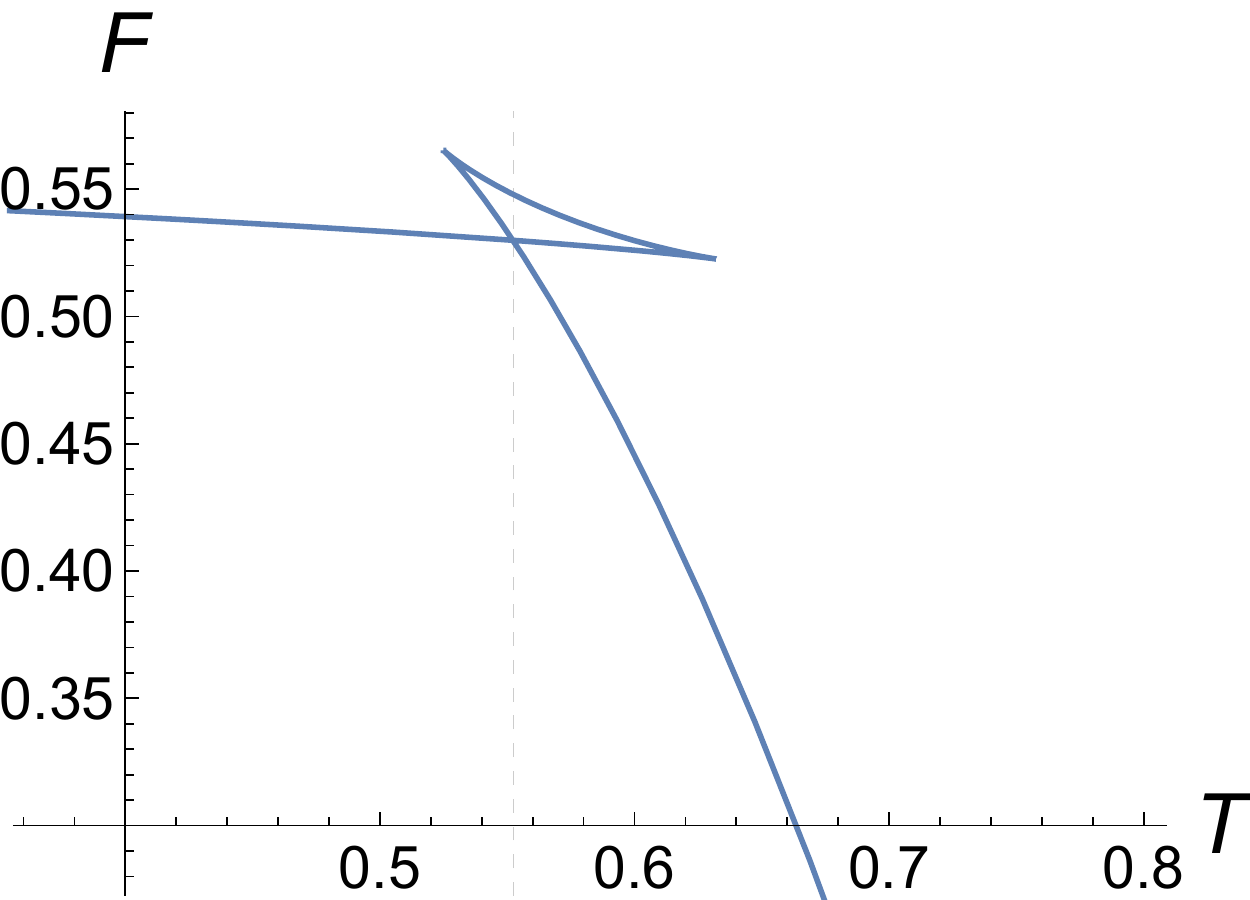}
  \end{center}
  \caption{Free energy plotted versus temperature for the following set of values: $J=-1$, $\pi^0=10$, $\pi^1=0.05$ (blue line, which correspond to the blue line of the previous plot as well). The derivative of the free energy is discontinuous for $T \simeq 0.55$, resulting in a first order phase transition: a small rotating black hole becomes a large one.\label{fe}}
\end{figure}

Determining the exact region of parameters $\{ \pi^0, \pi^1, J \}$ for the phase transition to arise is a daunting task given the complicated form of the solution. Of course, one can sample numerically the full region of parameters in the most general case, namely for all $\pi^0,\pi^1,J$ unconstrained. We refrain to do that but we provide some results in the simplified case when one of the charges is kept at a constant value. For instance, when we keep  $\pi^0 = 10$ fixed, there are maximal values $\pi^1_{max}$ and $J_{max}$ such that no phase transition can arise one of the charges, either $\pi^1$ or $J$, is greater than the respective maximal value:
\eq
\pi^1_{max} = 0.44 \,, \qquad  \quad J_{max} =3.75\,.
\feq
On the other hand, the phase transition occurs if both charges are lowered below these values:
\eq
\pi^1 < 0.33 \qquad and \qquad J < 2.55 \,,
\feq
while a case-by-case evaluation is needed for the intermediate values (e.g. $ 0.33 \leq \pi^1 \leq0.44$, $ 2.55 \leq J \leq 3.75 $) of the charges. 

The specific heat, computed as in \eqref{sph} is positive for large and small black holes, negative for the medium ones, and its behavior in function of the entropy is qualitatively similar to the one found in Fig. \ref{spec_heat}. 

\subsection{Dual field theory interpretation}

We would like now to interpret the phase transition we have found in the gravity side in the dual field theory. The latter belongs to the class of  ABJM models deformed by the insertion of operators with dimensions $\Delta=1,2$ and with background magnetic field. With reference to the analysis of the previous subsections, we consider the asymptotic behavior of each of the two scalar eigenmodes in eq. \eqref{as1} - \eqref{bc1} and \eqref{as2} - \eqref{bc2}. 

We now proceed to the study of the scalar  boundary conditions allowed in our system, in light of the results of \cite{Papadimitriou:2010as,Lindgren:2015lia}. There it is shown that holographic renormalization acts as a canonical transformation: the boundary counterterms necessary for a well defined variational principle (and for the removal of divergences) are such that the symplectic map $\phi*$ of the modes is diagonal. The symplectic map $\phi*$ (defined in Section 3 of \cite{Papadimitriou:2010as}), relates the variables $\varphi$, $\chi$ and the canonical momenta in the radial direction\begin{equation}
\pi_{\varphi} \equiv \sqrt{-g} \frac{\delta \mathcal{L}}{ \delta \partial_q \varphi} \,, \qquad \pi_{\chi} \equiv \sqrt{-g} \frac{\delta \mathcal{L}}{ \delta \partial_q \chi} \,, 
 \feq
to the space of asymptotic solutions, namely to the parameters we called $\alpha, \beta,\iota, \gamma$ in the previous sections. 

Given that the kinetic terms for the scalars in our case are non-canonical, one could expect mode mixing: all four parameters $\alpha, \beta, \iota, \gamma$ could in principle appear in $\pi_{\varphi}$ and $\pi_{\chi}$. In other words, the field/mode map might not diagonal. Since Dirichlet or Neumann boundary conditions must be imposed on the modes that make $\phi*$ diagonal, we need to make sure we are working with the correct eigenmodes.

In what follows we show that if we chose the parameterization \eqref{param_field}, $\phi*$ is indeed diagonal, hence we can correctly impose Dirichlet and Neumann boundary conditions on $\chi$ and $\varphi$.
 We work now in full generality, without reference to any particular solution of the equations of motion. We will use only the asymptotic expansion of the scalar fields $\varphi$ and $\chi$ (which we recall have mass $m^2 =-2$). 
 The asymptotic expansion at $r \rightarrow \infty$ of the fields and the related momenta reads:
\be
 \left( \begin{array}{c}  \pi_{\varphi} \\ \varphi 
\end{array} \right) = \left( \begin{array}{c}    -\frac{\alpha_{gen}}{2} q^2 - \beta_{gen} q + O(1)
 \\  \frac{\alpha_{gen}}{q} + \frac{\beta_{gen}}{q^2} + O(q^{-3})
\end{array} \right) \,.
\ee
For the sake of comparison, what we called here $\alpha_{gen}$ and $\beta_{gen}$ are identified with $\phi_{(0)}$ and $\phi_{(2\Delta-d)}$ of \cite{Papadimitriou:2010as}.
The field $\chi$ and its momentum read
\be
 \left( \begin{array}{c}  \pi_{\chi} \\ \chi 
\end{array} \right) = \left( \begin{array}{c}- \frac{ \iota_{gen}}{2} q^2 - \gamma_{gen} q + O(1)
 \\ \frac{\iota_{gen}}{q} +\frac{ \gamma_{gen}}{q^2} + O(q^{-3})
\end{array} \right) \,.
\ee
We see that $\pi_{\chi}$ depends on the two modes $\iota_{gen}, \gamma_{gen} \in \mathbb{R}$ and there is no mixing. 
We can then can use the procedure of \cite{Papadimitriou:2010as} (see also \cite{Lindgren:2015lia}) and add a canonical counterterm $S_{ct}$ \eqref{canon} which makes the symplectic map diagonal, giving 
\be
\phi* \left( \begin{array}{c}  \pi_{\varphi} \\ \varphi 
\end{array} \right) = \left( \begin{array}{c}  -\beta_{gen}  q + ...
 \\ \frac{\alpha_{gen}}{q} +...
\end{array} \right) \,,
\ee
and 
\be
\phi* \left( \begin{array}{c}  \pi_{\chi} \\ \chi 
\end{array} \right) = \left( \begin{array}{c}   -\gamma_{gen} q+...
 \\ \frac{\iota_{gen}}{q}+...
\end{array} \right) \,.
\ee
We can then impose Neumann BC setting to zero the second one of the modes, namely $\beta_{gen}=0$. We can decide moreover to impose $\iota_{gen}=0$ which enforces Dirichlet boundary conditions on $\chi$.

Comparing the general asymptotic expansion with that of our exact black hole solution we can read off the boundary conditions. Indeed in our case, due to  \eqref{as1}-\eqref{bc1}, we have $\beta_{gen}=0$, $\alpha_{gen} = \alpha = \sqrt2\Delta$ and due to  \eqref{as2}-\eqref{bc2}, $\iota_{gen}=0$, $\gamma_{gen} = \gamma =  \sqrt2\Delta j \cos \theta$.

Since the mass lies in the range \eqref{bfrange}, we are allowed to choose the alternative quantization for the field $\varphi$. Therefore $\alpha$ is expectation value of an operator $\mathcal{O}_1$ of dimension one. The source $\beta$ is set to zero. We can then consider $\alpha$ as order parameter of the phase transition, and we plot it as a function of the temperature. In analogy with \cite{Hristov:2013sya} there is a different behaviour in the two different regimes above the critical values of the charges and angular momentum, and below. The graphs resemble qualitatively those in Fig. \ref{cus2}, this we do not report them here.

The new ingredient of the analysis at this point is $\gamma$, asymptotic mode relative to the scalar $\chi$, corresponding to the expectation value of an operator $\mathcal{O}_2$ of dimension two.  We can make 3D plots of the value of the coefficient $\gamma$, in order to display its dependence on the temperature and $\theta$. As usual, we compare the two cases, the regime in which at least one of the charges $J, \pi^{\Lambda}$ is above its critical value (Fig. \ref{cup}) and that relative to charges smaller than their critical values (Fig. \ref{cup2}).

\begin{figure}[htb]
  \begin{center}
    \includegraphics[scale=0.7]{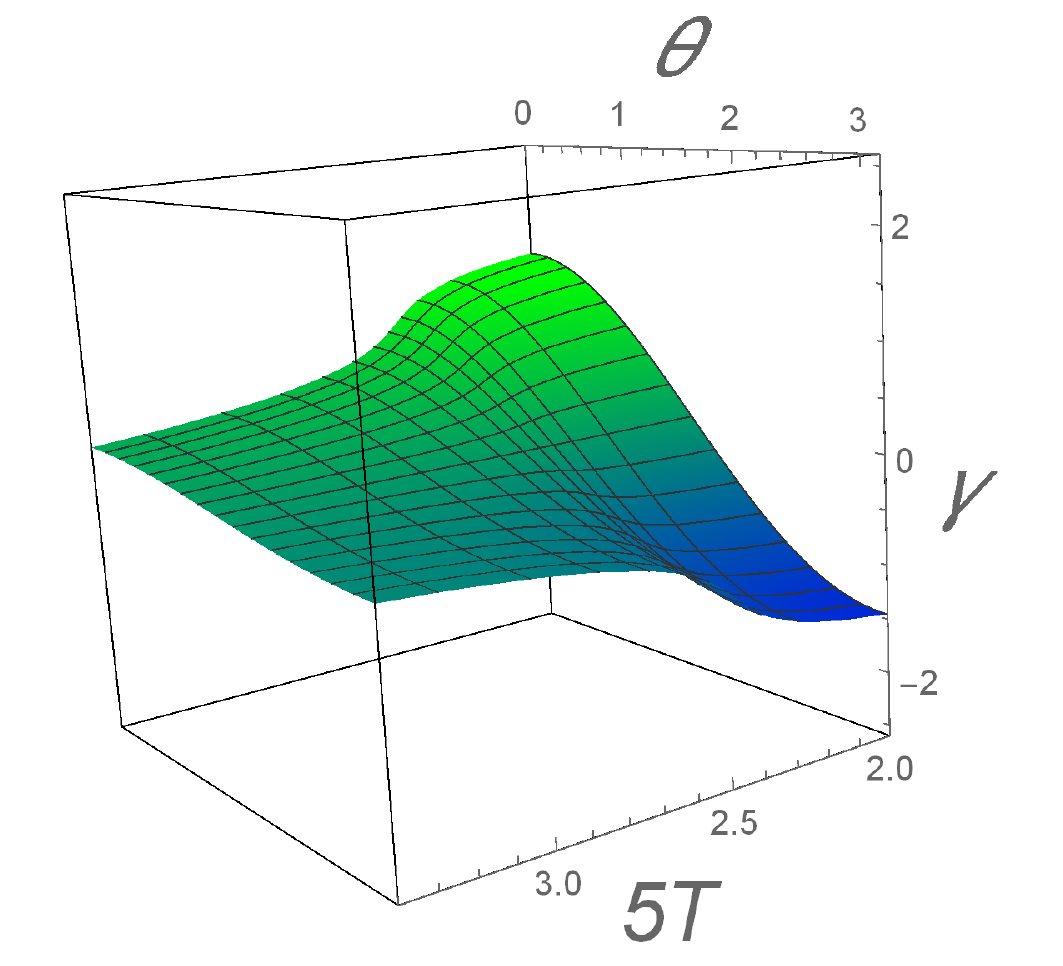}
  \end{center}
  \caption{Behaviour of $\gamma$ in function of temperature and angular variable $\theta$ in the regime where the phase transition does not happen: $\pi^0 = 6$, $\pi^1=0.5$, $J=0.1$. Indeed the temperature $T$ is always monotonic in function of $\gamma$.\label{cup}}
\end{figure}
\begin{figure}[htb]
  \begin{center}
    \includegraphics[scale=0.67]{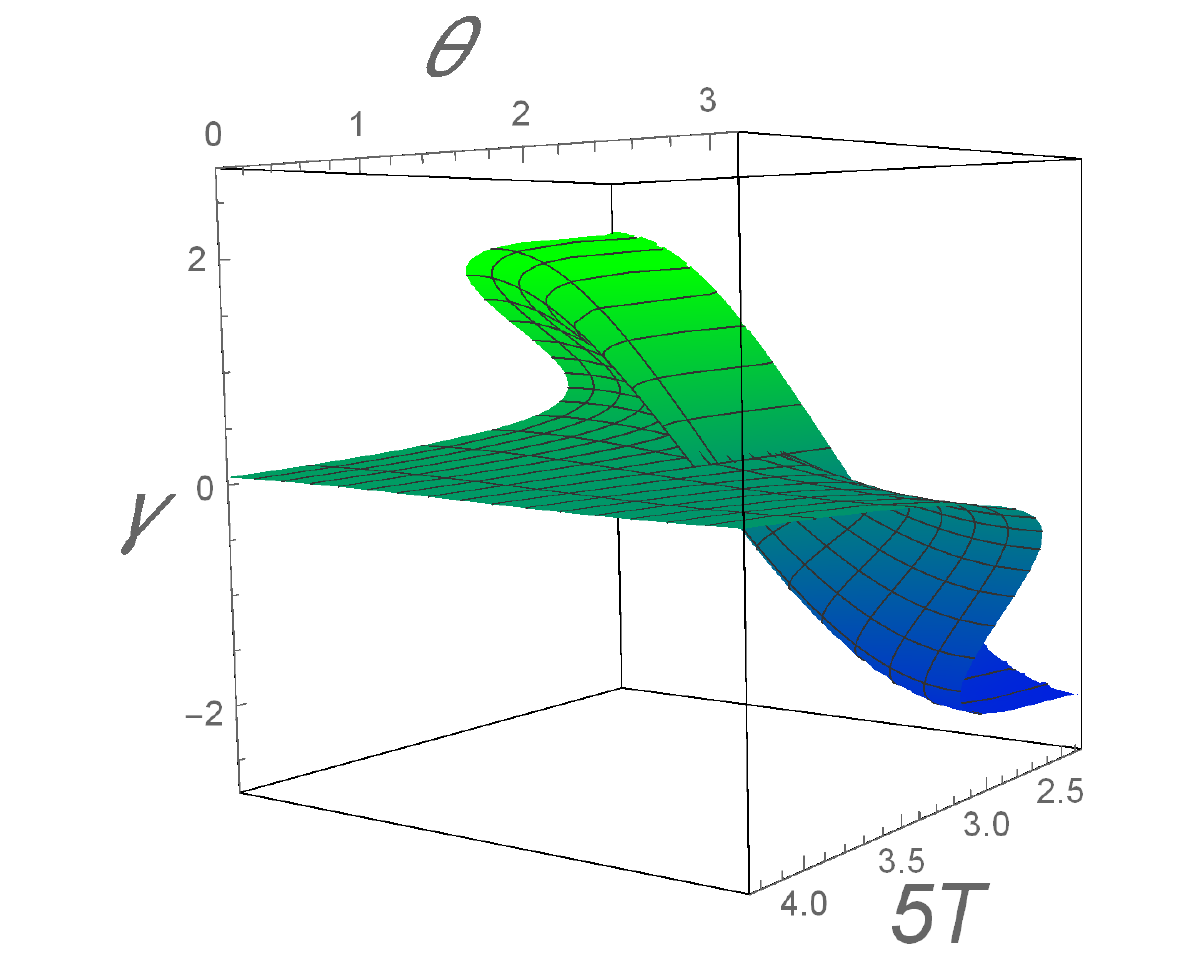}
  \end{center}
  \caption{Behaviour of $\gamma$ in function of temperature and angular variable $\theta$ in the regime $\pi^0=6$, $\pi^1=0.1$, $J=0.1$. We see that the order parameter exhibits a re-entrant behaviour in function of $T$, typical of liquid-gas like phase transitions. \label{cup2}}
\end{figure}

The behaviour of $\gamma$ in Fig. \ref{cup2} assumes a more complicated form. Nevertheless, for fixed $\theta$ slices the shape resembles the one of Fig. \ref{cus2}, namely it denotes a liquid-gas phase transition. Through the phase transition, the black hole hair decreases in absolute value in the southern hemisphere ($0< \theta < \pi/2$) while it increases in the northern one. That is a new feature of the phase transition that we have found here. Notice that there is no change in order parameter at the equator, where $\gamma=0$ always. For a positive value of the angular momentum $J$ the phase transition still happens, and the plot of $\gamma$ is reversed, namely the black hole hair increases in the southern hemisphere and it decreases in the northern one. Since $\gamma$ depends on the angular variable, it  breaks  rotational invariance in the boundary. A similar situation, where translational symmetry is spontaneously broken in the boundary due to spatially modulated instabilities in the bulk, is described for instance in \cite{Donos:2011qt}. 

It is worth noting at this point that, depending on boundary conditions satisfied by the scalar field, the phase transition manifest itself in the expectation value of different operators in the dual CFT, if we compare the cases $F = -2 \sqrt{X^0 (X^1)^3}$ (triple trace deformation) and $F= -i X^0 X^1$ (insertion of operators of dimension one and two). It would be interesting to find a deeper meaning of this phenomenon in the dual field theory. Of course, the fact that we are able to find analytically only configurations with certain boundary conditions for a given model might be due to our limitations in engineering a suitable ansatz\footnote{Among the limitations of our ansatz, for instance, a rotating solution with both non vanishing vector fields $A_0$ and $A_1$ has not yet been found for the square root prepotential.}. It would be interesting to see if a numerical investigation could unveil new solutions with different asymptotic scalar behaviour. Since the solutions treated here admit supersymmetric limits, we can are lead to think that the deformations of the ABJM models described here preserve supersymmetry. However, we cannot rule out the existence of more general black hole configurations which correspond to other deformations, which might break supersymmetry.

Before concluding this section, let us spend a few words about the singular limit $|j| \rightarrow l$. It was first shown in \cite{Gnecchi:2013mja,Klemm:2014rda} that this limit correspond to a configuration characterized by noncompact horizon of finite area. The event horizon in this case as a surface of revolution in $\mathbb{R}^3$ looks like an oblate sphere with two antipodal spikes, extending all the way to infinity (see Fig. 1 of \cite{Gnecchi:2013mja}). The entropy of the solution is finite and the temperature is as well. An interesting feature of this configuration is that it does not satisfy the Reverse Isoperimetric Inequality, therefore it corresponds to "super-entropic" states \cite{Hennigar:2014cfa}. The mass $M$ and the angular momentum $J$ are related by the chirality condition $M = -J/l^2$, which might be a hint of the underlying chiral microstate structure in the dual field theory (more details in \cite{Klemm:2014rda}). Regarding the study of the thermodynamics in the canonical ensemble, keeping both $J$ and $j$ fixed corresponds to choosing a particular value for $\Delta$. Hence for a given set of charges just one configuration exists and no phase transition can arise: it seems then that the angular momentum is too large for the solution to undergo such process.  It would be interesting to investigate if any instability arises for other choices of ensemble other than the canonical one.

\subsection{Addition of electric charges}

In this section we describe the black hole solution obtained by adding electric charges to the magnetic configuration (keeping the NUT charge to zero). It has this form
\begin{eqnarray}\label{met2} \nonumber
ds^2 &= & -\frac{Q(q)}{(q^2-\Delta^2-d^2+j^2\cos^2\!\theta)}\left[dt + \frac{j\sin^2\!\theta}{\Xi} d\phi\right]^2 \\
& +& (q^2-\Delta^2 -d^2+j^2\cos^2\!\theta)\left(\frac{dq^2}{Q(q)} + \frac{d\theta^2}{\Delta_{\theta}}\right) \nonumber \\
             &+ &\, \frac{\Delta_{\theta}\sin^2\!\theta}{(q^2-\Delta^2-d^2+j^2\cos^2\!\theta)}\bigg[ \,\, j \, dt + \nonumber \\
             & & \qquad \qquad  \qquad+
                   \frac{q^2+j^2-\Delta^2-d^2}{\Xi}d\phi\bigg]^2\,, \label{metr-nonextr-spher-X0X1} 
\end{eqnarray}
with $\Delta_{\theta}$  and $\Xi$ defined as in \eqref{deltatheta}. 
The vectors fields are given by
\eq
A^\Lambda = \frac{\mathsf{P}^\Lambda(dt + q_1q_2dy)}W p_1 + \frac{\mathsf{Q}^\Lambda(dt - p_1p_2dy)}{W} q_1 \,,
\feq
where 
\eq \nonumber
W = q_1q_2 + p_1 p_2 \,,  \quad q_1 = q-\Delta\,, \quad q_2=q+\Delta\,,
\feq 
\eq 
 p_1 = p-d\,, \quad p_2=p+d\,, \quad p=j \cos \theta\,.
\feq
The function $Q(q)$ is defined as in \eqref{QP} and contains the parameters:
\begin{eqnarray} \nonumber
a_0 & = & (j^2-\Delta^2-d^2)\left(1-\frac{\Delta^2+d^2}{l^2}\right) + \\ \nonumber
&  &  + \,  2 \, l^2\left(g_0^2({\mathsf{P}^0}^2 + {\mathsf{Q}^0}^2 )+
g_1^2({\mathsf{P}^1}^2+{\mathsf{Q}^1}^2 ) \right)\,, 
\end{eqnarray}
\eq \nonumber
 a_1= \frac{2 l^2 (g_0^2 ({\mathsf{P}^0}^2 - {\mathsf{Q}^0}^2 ) - g_1^2({\mathsf{P}^1}^2-{\mathsf{Q}^1}^2)}
{ \Delta}\,, 
\feq
\eq
a_2 = 1-\frac{\Delta^2+d^2}{l^2} + \frac{j^2-\Delta^2-d^2}{l^2}\,, 
\nonumber
\feq
\eq
 a_4 = 1/l^2 \equiv 4g_0g_1 \,.
\feq
Finally, $d$ is given by:
\eq
d = -\frac{2 \Delta \left(g_0^2 \mathsf{P}^0\mathsf{Q}^0-g_1^2 \mathsf{P}^1 \mathsf{Q}^1 \right)}{g_0^2 ({\mathsf{P}^0}^2- {\mathsf{Q}^0}^2 )-g_1^2 ( {\mathsf{P}^1}^2- {\mathsf{Q}^1}^2)}\,.
\feq
This configuration corresponds to the solution of \cite{Chow:2013gba} obtained by setting the NUT charge parameter to zero. In total the configuration is parameterized by 8 real parameters, $g_0$, $g_1$, $\mathsf{P}^0$, $\mathsf{P}^1$, $\mathsf{Q}^0$ , $\mathsf{Q}^1$, $\Delta$ and $j$, corresponding to the FI terms (that determine the value of the cosmological constant), 4 electromagnetic charges, mass and angular momentum.

In this case the scalar field has the following asymptotic behaviour:
\eq
\varphi = \frac{\alpha}{q} + \frac{\beta}{q^2} + O(r^{-3})\,,
\feq
where 
\eq
\alpha =  \sqrt2 \Delta \,,  \qquad \beta  =   \sqrt2 d p
\feq
and 
\eq \label{expchi}
\chi =  \frac{\iota}{q} + \frac{  \gamma}{q^2} + O(q^{-3})\,, 
\feq
where
\eq
\iota=  \sqrt2 d \,,
\qquad
\gamma= \sqrt2 \Delta p \,.
\feq

We see from here that the dependence on the angular variable $p$ is already present in the $q^{-2}$ term of the eigenmode $\varphi$. The presence of dyonic charges manifests itself also in the presence of a nonzero term $\iota$ in the expansion of the second scalar eigenmode $\chi$.

Notice that, in particular, taking again $g_0=g_1=1/2$, switching off either both electric  charges $\mathsf{Q}^0=\mathsf{Q}^1=0$ or switching off both magnetic charges $\mathsf{P}^0=\mathsf{P}^1=0$ enforces $\iota= \beta=0$, which is the case studied previously. The same scalar boundary conditions are found here also by setting
\be \label{relaz}
\mathsf{P}^0 \mathsf{Q}^0  = \mathsf{P}^1 \mathsf{Q}^1 \,.
\ee
To sum up, either having charges of the same nature (either all electric, or all magnetic in the symplectic frame considered) or charges that satisfy \eqref{relaz} are the conditions leading to boundary conditions \eqref{bc1}-\eqref{bc2} for the scalar fields. We can impose then Dirichlet boundary conditions on $\chi$ and Neumann on $\varphi$, as before, so that the black hole configurations can be compared as states in the same theory with zero source.  We have verified that for all these cases the $T-S$ and $F-T$ plots assume the same qualitative for as Figs. \ref{cup0}-\ref{fe}, hence a phase transition takes place.

The allowed boundary conditions for the thermodynamic quantities like the mass to be well defined were studied in \cite{Chow:2013gba}. There it was found that for this particular model the symplectic structure is conserved, hence a hamiltonian is well defined, for all sets of charges (while examples of models where this is not true can be found in \cite{Lu:2013ura}).
The explicit computation of the on-shell action and mass via holographic renormalization in the simple example of purely magnetic configuration is treated in appendix.

\section{Acknowledgements}

We would like to thank A. Amariti, F. Denef, D. Klemm, S. Vandoren for interesting discussions, C. Asplund, K. Hristov  and R. Monten for carefully reading the manuscript and in particular Ioannis Papadimitriou for stimulating discussions and important clarifications. We furthermore thank the referees for the useful comments. We are grateful to the organizers and the participants of the of the conference "Theoretical frontiers in Black holes and Cosmology" for interesting discussions during the completion of this work. We acknowledge support from the NWO Rubicon grant, Columbia University and from DOE grant DE-SC0011941.

\section{Appendix: Mass and on-shell action from holographic renormalization} \label{mass}

Given the boundary conditions that we have found in the previous sections, we can compute the on shell action and conserved charges for the black hole configuration via holographic renormalization. We follow the conventions of \cite{Papadimitriou:2007sj}, where the necessary counterterms were identified. After describing the general form of the canonical counterterms, we spell out in detail the computation for the $F = -i X^0 X^1$.

We start from the action $S = S_{bulk} + S_{bdary}$, where
\begin{eqnarray} \nonumber
S_{bulk} &=& -\frac{1}{16 \pi } \int_M  d^4x \sqrt{-g} \, \bigg[ \frac{R}{2} -g_{\tau \bar{\tau}} \partial_{\mu} \tau \partial^{\mu} \bar{\tau} -V_g \\
&& +  I_{\Lambda \Sigma} F^{\Lambda}_{\rho \sigma} F^{\Sigma , \rho \sigma} + \frac12 R_{\Lambda \Sigma} \epsilon^{\mu \nu \rho \sigma} F_{\mu \nu}^{\Lambda} F_{\rho \sigma}^{\Sigma} \bigg] \nonumber \\
S_{bdary} &= &- \frac{1}{8 \pi} \int_{\partial M} d^3x \sqrt{-h} \Theta -\frac{1}{16 \pi}S_{ct}\,. \nonumber \\
\end{eqnarray}
The last two terms are function of the boundary 3d metric $h_{ab}$, and consist of the Gibbons-Hawking boundary term and the canonical counterterm $S_{ct}$\footnote{We must add the term imposing the Neumann boundary conditions on $\varphi$ as well. The form of this term can be read in \cite{Papadimitriou:2007sj}, Table 2, term $S_{-}$, however this has a vanishing contribution on the on-shell action for these solutions, hence we do not report it here.}.

The Gibbons-Hawking term is a function of the extrinsic curvature $\Theta_{\mu \nu}$ computed as
\eq
\Theta= \Theta_{\mu}^{\mu} \,, \qquad \Theta_{\mu \nu} = - (\nabla_{\mu} n_{\nu} + \nabla_{\nu} n_{\mu}) \,,
\feq
where $n_{\mu}$ is the unit vector normal to the boundary (which in our case is located at $q \rightarrow \infty$)\footnote{Alternative but equivalent renormalization procedures involve the addition of topological invariants such as the Gauss--Bonnet term, see i.e. \cite{Olea:2005gb,Miskovic:2009bm}.}.
The term $S_{ct}$ is the ordinary counterterm obtained via the Hamilton-Jacobi procedure:
\begin{equation} \label{canon}
S_{ct}=\int_{\partial M} d^3x \sqrt{-h} \left[\frac{l}{2} \mathcal{R} - \frac{l^3}{2} \left( \mathcal{R}_{bc} \mathcal{R}^{bc} -\frac{3 \mathcal{R}^2}{8}  \right) +\mathcal{W}  \right]\,,
\end{equation}
where the quantities $h_{ab}$ and $\mathcal{R}_{ab}$ denote the metric and Ricci curvature of the boundary $\partial M$. Once again we set $l=1$.
The superpotential $\mathcal{W}$ appearing in \eqref{canon} satisfies this relation:
\eq \label{fundrelationpotential}
V_g = \frac12 \left(-\frac32 \mathcal{W}^2 + g^{ij} \partial_i \mathcal{W} \partial_j \mathcal{W} \right)\,,
\feq
where $i,j$ are indices referring to the real scalar fields of the theory under consideration ($\varphi$ and $\chi$, in our case). This equation can be expanded around in double series in $\varphi$ and $\chi$ around zero, and by coefficient matching one can read off the expansion of the superpotential. From now on we report the explicit computation for the solution of the model $F= -i X^0 X^1$. The expansion of the superpotential reads:
\eq
\mathcal{W}_{\pm}= -2 -\frac12 \Delta_{\pm} (\chi^2 + \varphi^2) + O(\varphi^3, \chi^3) \,.
\feq
The two choices in $\mathcal{W}$ are related to the dimensions  $\Delta_{\pm}$ of the operators dual to the corresponding fields of mass $m^2=-2$, hence $
\Delta_+ =2$, $\Delta_{-} =1$,
as was noticed for example in  \cite{de Boer:1999xf}. The holographic renormalization procedure prescribes the use of the superpotential $\mathcal{W}_{-}$  as canonical counterterm \cite{Papadimitriou:2006dr,Papadimitriou:2007sj}. Indeed, only a superpotential of this form is suitable for removing the divergences from the action. Therefore when using formula \eqref{canon} we will take $\mathcal{W} = \mathcal{W}_{-}$. We are going to compute first the free energy, then mass and angular momentum, and we express the renormalized on-shell action in terms of the other conserved quantities \eqref{helm}.

The use of the Einstein's equations of motion allows us to rewrite the bulk on shell action as
\begin{eqnarray} \label{bulk}
S_{bulk} = -\frac{1}{16 \pi }& \int_M&  d^4x \sqrt{-g} \, \bigg[ V_g+  \\
&+&  I_{\Lambda \Sigma} F^{\Lambda}_{\rho \sigma} F^{\Sigma , \rho \sigma} + \frac12R_{\Lambda \Sigma} \epsilon^{\mu \nu \rho \sigma} F_{\mu \nu}^{\Lambda} F_{\rho \sigma}^{\Sigma} \bigg]\,. \nonumber 
\end{eqnarray}
We regularize at the boundary, by imposing a radial cutoff $q_0$. The integration on the radial variable $q$ is then performed with extrema $q_0$ and $q_h$, radius of the event horizon. Integration of the first term of \eqref{bulk} gives a contribution of
\begin{eqnarray}
 -\frac{1}{16 \pi }\int_M  d^4x  \sqrt{-g} \, V_g = &-&\frac{\beta_t}{ 4\Xi} \big[ q_0 \left( j^2+ q_0^2 -  \Delta^2 \right)+  \nonumber \\
 &-&   q_h \left( j^2+ q_h^2 -\Delta^2\right) \big]\,,
 \end{eqnarray}
 where $\beta_t$ arise from the integration over the time coordinate. The second term gives a finite term which is
\begin{eqnarray}
-\frac{1}{16 \pi } & \int_M & d^4x \sqrt{-g}  \left[ I_{\Lambda \Sigma} F^{\Lambda}_{\rho \sigma} F^{\Sigma , \rho \sigma} + \frac12 R_{\Lambda \Sigma} \epsilon^{\mu \nu \rho \sigma} F_{\mu \nu}^{\Lambda} F_{\rho \sigma}^{\Sigma} \right] \nonumber \\
& = &  \beta_t \frac{\Delta ((\mathsf{P}^0)^2-(\mathsf{P}^1)^2) +q_h \left((\mathsf{P}^0)^2+(\mathsf{P}^1)^2\right)}{ 4\Xi \left(\Delta^2-j^2-q_h^2\right)} \,.\nonumber \\
\end{eqnarray}

The counterterm action gives the following contribution:
\eq
S_{ct} = \frac{\beta_t}{ 4\Xi} \left( 2 q_0 \left(2 j^2+2 q_0^2\right) + a_1\right)\,,
\feq
hence the total action is free from divergences and it assumes the following form:
\eq \nonumber
\frac{S_{onshell}}{\beta_t} =
-\frac{a_1}{4 \Xi} +\frac{ \Delta ((\mathsf{P}^0)^2-(\mathsf{P}^1)^2) +q_h \left((\mathsf{P}^0)^2+(\mathsf{P}^1)^2\right)}{(-\Delta^2+j^2+q_h^2) 4 \Xi} 
\feq
\eq
-\frac{q_h \left(-\Delta^2+j^2+q_h^2\right)}{2\Xi}\,.
\feq
We now turn to the computation of the  conserved charges, which are extracted from the stress-energy tensor $\tau^{ab}$
\begin{equation}
\tau^{ab} = \frac{2}{\sqrt{-h} } \frac{ \delta S}{ \delta h_{ab}}\,.
\end{equation}
For any Killing vector field $\xi^a$ associated with an isometry of the boundary 3-metric, we can define the conserved quantity
\begin{equation}\label{charge}
Q_{\xi} = \frac{1}{8 \pi} \int_{\Sigma} \sqrt{\sigma} u_{a} \tau^{ab} \xi_{b}\,,
\end{equation}
where $\Sigma$ is the spacelike section of the boundary metric.
The computation of $\tau^{ab} $ gives\begin{eqnarray}
\tau^{ab} & =&  -\frac{1}{8 \pi} \bigg[ (\Theta^{ab} -h^{ab} \Theta) + \mathcal{W} h^{ab} \nonumber \\
&- & \big(\mathcal{R}^{ab} -\frac12 h^{ab} \mathcal{R} \big) \bigg]\,,
\end{eqnarray}
where we neglected terms quadratic in $\mathcal{R}_{ab}$ because subleading when $q \rightarrow \infty$.
Using \eqref{charge} the renormalized stress-energy tensor turns out to be
\eq
\tau_{tt} = -\frac{a_1}{ q} +O\left( \frac{1}{q^2}\right)\,,
\feq
hence the mass is
\eq
M = Q_{\partial_t/\Xi} = -a_1/ (2 \Xi^2)\,,
\feq
in agreement with the result of the AMD procedure. Given the expressions for the other conserved charges \eqref{magnch}, potentials \eqref{magn-pot}, \eqref{ome} and thermodynamics variables \eqref{entropy}  we see that the on shell action coincides with 
\eq
\frac{S_{onshell}}{\beta_t} = M- TS -\Omega J\,,
\feq
as expected for magnetic configurations, see for instance the discussion in \cite{Hawking:1995ap}. We perform a Legendre transform with repect to $\Omega$ in order to keep the angular momentum fixed, obtaining the Helmholtz free energy  \eqref{helm} used in the thermodynamics analysis.

As a last remark, we would like to point out that for our purposes if was sufficient to know the expansion to second order of the superpotential $\mathcal{W}$. Alternatively, we could have used the exact superpotential driving the flow of the scalar field in the static case, in this way $\varphi' = \partial_{\varphi} \mathcal{W}_{flow}$. The latter is known for the static $J=0$ solution \cite{Cacciatori:2009iz,Klemm:2012yg}:
\eq \label{supreal}
\mathcal{W}_{J=0} = - e^{-\varphi / \sqrt2} - e^{\varphi/ \sqrt2} = -2 \cosh \left(\frac{\varphi}{\sqrt2} \right)\,.
\feq 
Obviously the expansion at second order of \eqref{supreal} gives exactly $\mathcal{W}_{11}$, so the use of this counterterm leads to the same conserved charges.

The holographic renormalization for the electric configurations proceed in the same way as we just did, except for the fact that the presence of mixed boundary condition on the scalar field prescribes the addition of a finite counterterm, along the lines of \cite{Papadimitriou:2007sj,Gnecchi:2014cqa}. Such counterterm can be reabsorbed in the canonical counterterms if we use the correct superpotential, namely the one that correctly drives the flow of the scalar fields in the static case. Such superpotential is:
\eq
 \mathcal{W} = e^{-\sqrt{3/2}\varphi}+3 e^{\varphi/\sqrt6} \,.
\feq
We do not report the full computation here, but further details about the procedure for static black holes can be found in \cite{Batrachenko:2004fd, Gnecchi:2014cqa}.


\end{document}